# Mobile XR over 5G: A way forward with mmWaves and Edge

*Cristina Perfecto*[1], *Mohammed S. Elbamby*[2], *Jihong Park*[2], *Javier Del Ser*[1,3,4] *and Mehdi Bennis*[2]
[1]*University of the Basque Country (UPV/EHU), Spain.*
[2]*Centre for Wireless Communications, University of Oulu, Oulu, Finland.*
[3]*Tecnalia Research & Innovation, Bilbao, Spain.*
[4]*Basque Center for Applied Mathematics (BCAM), Bilbao, Spain.*
[cristina.perfecto@ehu.eus](cristina.perfecto@ehu.eus), [mohammed.elbamby@oulu.fi](mohammed.elbamby@oulu.fi), [jihong.park@oulu.fi](jihong.park@oulu.fi),
[javier.delser@tecnalia.com](javier.delser@tecnalia.com), [mehdi.bennis@oulu.fi](mehdi.bennis@oulu.fi)

## 1. Introduction

After having been labeled as a gamers' and geeks' technology and set aside from the mainstream consumer market, extended reality (XR) has re-emerged fueled by the promise of a mobile interconnected VR [1] and of a future Tactile Internet (TI) [2], that is called to allow remote interaction with real and virtual elements (objects or systems) in perceived real-time[1].

Accordingly, the anticipated application portfolio for XR –a term that encompasses all virtual or combined real-virtual environment compounds including virtual reality (VR), augmented reality (AR) and mixed reality (MR)– spans beyond immersive live-sport retransmissions, gaming or 360º video and finds its natural soil in areas of robotics for health care and smart factory environments with remote surgery their best representative. All these applications will require different levels of immersion built on extremely high-quality multi-sensory XR experiences; even more notably so those servicing critical areas for society [3].

However, immersion is fragile and sustaining it along time is computationally intensive which, paired to its acute sensitiveness to delay, makes it easy to break. Focusing on the visual response, there is a broad consensus that an end-to-end (E2E) delay, also known as motion-to-photon (MTP) delay, of 10-20 milliseconds is the maximum allowable in XR. Exceeding these values causes a visual-motor sensory conflict that might eventually trigger an episode of motion sickness. Likewise, to provide human tactile to visual feedback control, round-trip latencies below 1 ms together with robustness and availability will be needed.

Breakthroughs in computing and communication need to be harnessed to reduce latency, enhance reliability, and improve scalability, such that perceived real-time operation in multi-sensory XR, as the forerunner for TI and haptic communications [4][5], becomes feasible under resource constraints and the uncertainty arriving from wireless channel dynamics. In this regard, this e-letter summarizes our recent work and proposed approaches in [6]–[9] that weave together several technologies from emerging 5G communications systems towards enabling a fully immersive experience.

## 2. XR Requirement Triangle: Capacity, Latency, and Reliability in Scale

From the original three service categories in the fifth generation (5G), enhanced mobile broadband (eMBB) has been so far the one where most progress has been made towards boosting the capacity and enhancing connectivity to attain the anticipated 10/20 Gbps of peak data rate in UL/DL [10]. As opposed to these advancements, the utterly different statistical treatment required by principled ultra-reliable and low-latency communication (URLLC) [11] has hampered its progression that is still at its infancy. Mobile multi-sensory XR sits somewhere in between eMBB and URLLC requiring data rates covering from 100 Mbps (1K entry-level VR resolution) to up 1 Gbps (compressed human eye resolution) delivered uniformly [6] to end-users. Moreover, this delivery is further subject to latency constraints that significantly differ based on whether an exclusively visual or a visual plus haptic response is sought. As these values are clearly currently unrealizable, we postulate that smart network designs combining the use of higher frequency bands, multi-access edge computing (MEC) [12], edge machine learning (edgeML) [13], and decision making frameworks that incorporate the notions of risk and tail distributions of extreme events will need to be recruited.

Next, we summarize some of the approaches contributed in our latest works and outlined in Figure 1, to increase the capacity, cut down on the latency and provide higher reliability in several VR scenarios.

---

[1] Perceived real-time implies that the latency incurred by computing and communication is negligible for the considered sensory context (muscular, audio, visual, and haptic or tactile) whereby human reaction and interaction times range from 1 s to 1 ms.

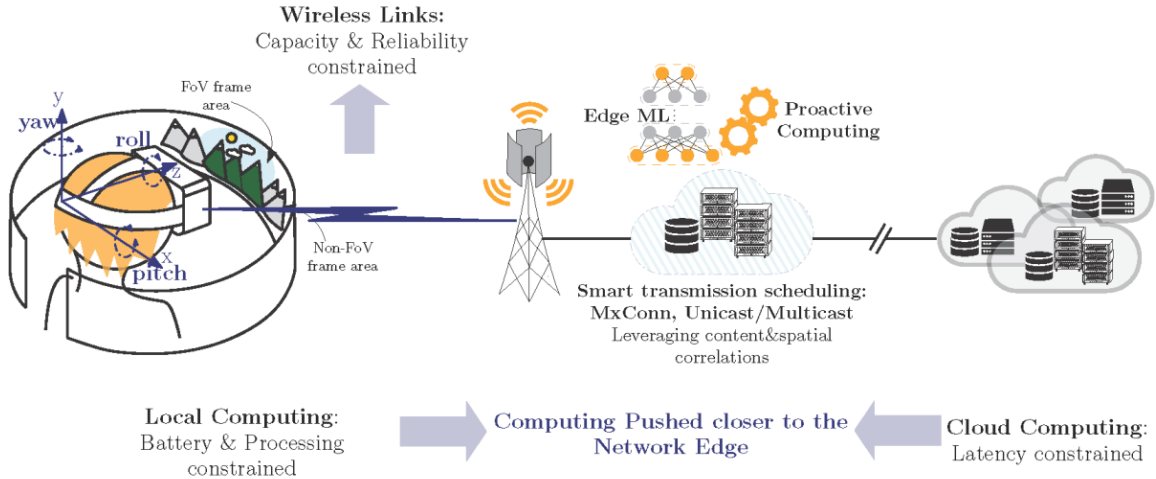

Figure 1: Illustration of some of the communication and computing related bottlenecks and possible enablers to provide a high-quality mobile XR experience.

### 2.1. Confronting the capacity crunch

Given the current spectrum shortage in the sub 6GHz cellular bands, the use of the millimeter wave (mmWave) frequency bands [14] seems the natural remedy to solve the clash between the available vs. demanded bandwidth for a UHD quality wireless VR [6]. In this regard, if the transmitter and receiver are in line-of-sight (LOS) and the mainlobes of their respective antenna radiation patterns pointing towards each other, i.e. aligned, the use of mmWave communications grants an immediate channel capacity increase. Nevertheless, mmWave links are vulnerable to blockage and beam misalignment. Hence, the resulting channel can be highly intermittent. For this reason, we strongly advocate for the use of mmWaves complemented with techniques to enhance the reliability of wireless links as exemplified in [7] and further discussed in Subsection 0.

An altogether different approach that has lately concentrated a significant amount of research efforts is aimed to reduce the bandwidth needs in mobile/wireless VR, thereby shrinking the amount of data processed and transmitted. In field-of-view adaptive streaming (FOVAS), raw 360º immersive VR video frames are spatially segmented[2], and only those portions that fall within the field of view (FoV) are transmitted in HD. For that purpose, the head movements need to be tracked and then, adopting a tiled[15] or viewport [16] based frame decomposition, decide on the parts of the video frame to be transmitted either real-time or based on estimations supplied by a companion machine learning (ML) backend. These latter predictions also allow carrying out proactive content transmissions.

### 2.2. Taming the latency

There are manifold aspects that contribute to E2E latency for XR. In this regard, enabling low latency requires intertwining several techniques to be implemented both at the computing/processing and communication levels.

Firstly, due to wireless VR headset limited computing resources, it is mandatory to offload computing-intensive tasks to servers and adopt the proximity computing commissioned by MEC whereby computing, content, and connectivity services are pushed closer to the data source and consumption points.

Secondly, considering the communication level, exploiting proximity computing and mmWave links play a significant role in reducing the latency as 1) the distance between the end devices and the edge servers is shrunken and 2) efficient wireless/wired backhauling with low latency access to MEC services is achieved.

At the computing processing level and related to proactive content caching, where knowledge of users' preferences and future interests allow for prefetching of their content, data availability and edgeML will help to speed up computing the tasks of network nodes. The latter idea is exemplified for XR in [8], [17] and [18] to predict users' future FoV and subsequently empowering data correlation to reduce latency and resource utilization.

---

[2] An alternative rendering method, foveated rendering, integrates eye gaze tracking in the VR headset and transmits high-resolution content only for the areas of the frame that correspond to the center of the human vision (fovea centralis) while greatly reducing resolution and color depth in the peripheral field of view.

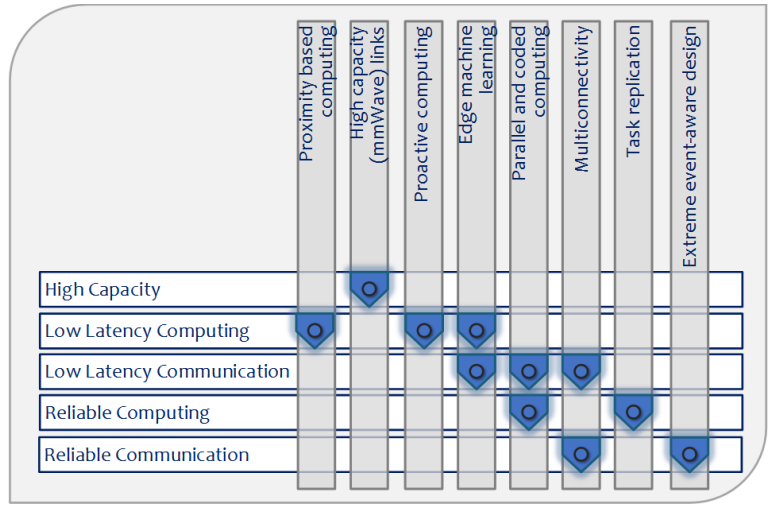

Figure 2: Mapping matrix of technological enablers to reduce latency and improve reliability in communication and computing related to mobile XR.

### 2.3. Enhancing Reliability

Reliability refers to the availability/provisioning of a certain level of communication or computing service with some assured guarantees, e.g., 99.99999 percent of the time. Nonetheless, a second interpretation that is widely adopted among wireless communications standardization bodies treats reliability as a probabilistic bound over the maximum allowable latency, i.e., it is interpreted as a delay-reliability.

No matter in which of its connotations, reliability is a crucial element in future XR applications. In what relates to reliability in its original meaning, the adoption of mmWave links to deliver the visual traffic required high data rates comes at the cost of dealing with a more vulnerable channel, mainly due to signal blockage. A more robust mmWave communication is achievable by embracing multi-connectivity (MC). MC encompasses several techniques developed to increase effective data rates and mobility robustness of wireless links. For that purpose, diversity is applied to cut down on the number of failed handovers, dropped connections and radio-link failures (RLFs) originated service interruptions. MC allows users to establish simultaneous connections with multiple base stations (BSs) in the same frequency channel, i.e., intra-frequency MC, or through different channels/interfaces, i.e., inter-frequency MC.

As for delay/latency related reliability, which is more concerned with the reducing delay tail rather than with the average delay, low-latency enablers, such as proactive computing, can be useful. It should be noted here, that there is a clear tradeoff between minimizing the latency in general and providing guarantees on the delay exceedance. Therefore, it is essential to design XR systems with tools that look into characterizing the extremely rare conditions of delay, such as extreme-value theory (EVT) [19] [20].

### 3. Resource Provisioning for Multi-modal sensory information with EMBB/URLLC slicing

To realize a multi-sensory XR, flexible approaches to radio resource management, capable of providing on-demand functionality, would be essential in 5G networks. For instance, a key challenge for multi-sensory XR arises from having different sensory contexts with different requirements in terms of sampling, transmission rate, and latency which is usually referred to as *cross-modal asynchrony*. Accordingly, visuo-haptic XR traffic entails the use of two different network slices: eMBB for visual perception and URLLC for haptic perception. Therefore, a multiplexing scheme is required that is capable of exploiting priorities as well as temporal integration of these different modalities. Our work in [9] investigates how to share the DL resources orthogonally and non-orthogonally, respectively in terms of the impact in the just-noticeable difference (JND), a measure to describe the minimum detectable change amount of perceptual inputs, of the aggregate visuo-haptic perception.

As URLLC traffic cannot be queued due to its hard latency requirements, radio resources must be provided with priority for haptic communications. To that end, URLLC traffic is usually scheduled on top of the ongoing, i.e., puncturing, eMBB transmissions. Our work in [21] applies a risk-sensitive formulation to allocate resources to the incoming URLLC traffic while minimizing the risk of the eMBB transmission. Thereby low data rate eMBB users are protected while ensuring URLLC.

## 4. Conclusion

This letter has presented a summary of our latest research efforts towards providing an immersive wireless XR experience as a first stepping stone to wirelessly delivering multi-sensory XR experiences over 5G networks.

**Acknowledgment**

This work was supported in part by the Academy of Finland project CARMA, in part by the Academy of Finland project MISSION, in part by the Academy of Finland project SMARTER, in part by the INFOTECH project NOOR and in part by the Spanish Ministerio de Economía y Competitividad (MINECO) under grant TEC2016-80090-C2-2-R (5RANVIR).

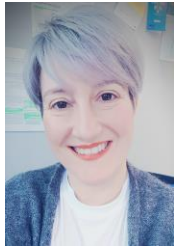

**Cristina Perfecto** (S'15) received her M.Sc. degree in Telecommunication Engineering by the University of the Basque Country (UPV/EHU), Spain, in 2000. She is currently a College Associate Professor at the Department of Communications Engineering of this same University. Her research interests are machine learning and data analytics including different fields such as metaheuristics and bio-inspired computation, both from a theoretical and applied point of view. She is currently working towards her Ph.D. focused on the application of multidisciplinary computational intelligence techniques in radio resource management for 5G, specifically on resource optimization for V2X and XR operating mmWave communications.

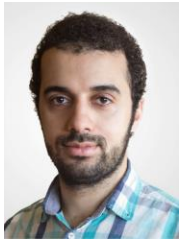

**Mohammed S. Elbamby** (S'14) received the B.Sc. degree (Hons.) in Electronics and Communications Engineering from the Institute of Aviation Engineering and Technology, Egypt, in 2010, and the M.Sc. degree in Communications Engineering from Cairo University, Egypt, in 2013. He is currently pursuing the Dr.Tech. Degree with the University of Oulu. After receiving the M.Sc. degree, he joined the Centre for Wireless Communications, University of Oulu. His research interests include resource optimization, uplink and downlink configuration, fog networking, and caching in wireless cellular networks. He received the Best Student Paper Award from the European Conference on Networks and Communications in 2017.

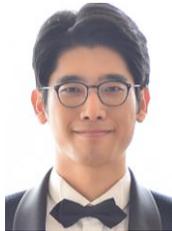

**Jihong Park** received the B.S. and Ph.D. degrees from Yonsei University, Seoul, South Korea, in 2009 and 2016, respectively. From 2016 to 2017, he was a Post-Doctoral Researcher with Aalborg University, Denmark. Dr. Park was a Visiting Researcher with Hong Kong Polytechnic University; KTH, Sweden; Aalborg University, Denmark; and New Jersey Institute of Technology, USA, in 2013, 2015, 2016, and 2017, respectively. He is currently a Post-Doctoral Researcher with the University of Oulu, Finland. His research interests include ultra-dense/ultra-reliable/massive-MIMO system designs using stochastic geometry and network economics. His papers on tractable ultra-dense network analysis received the IEEE GLOBECOM Student Travel Grant in 2014, the IEEE Seoul Section Student Paper Contest Bronze Prize in 2014, and the 6th IDIS-ETNEWS (The Electronic Times) Paper Contest Award sponsored by the Ministry of Science, ICT, and Future Planning of Korea.

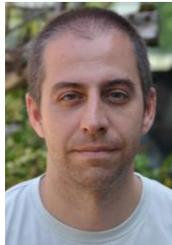

**Javier del Ser** (SM'14) received his first Ph.D. in Telecommunication Engineering (Cum Laude) from the University of Navarra, Spain, in 2006, and a second Ph.D. in Computational Intelligence (Summa Cum Laude) from the University of Alcala, Spain, in 2013. He is currently a principal researcher in data analytics and optimization at TECNALIA (Spain), an associate researcher at the Basque Center for Applied Mathematics and an adjunct professor at the University of the Basque Country (UPV/EHU). His research activity gravitates on the use of descriptive, prescriptive and predictive algorithms for data mining and optimization in a diverse range of application fields such as Energy, Transport, Telecommunications, Health, and Security, among many others. In these fields, he has published more than 160 publications, co-supervised 6 Ph.D. theses, edited three books, co-authored six patents and led more than 35 research projects. Dr. Del Ser has been awarded the *Talent of Bizkaia* prize for his curriculum.

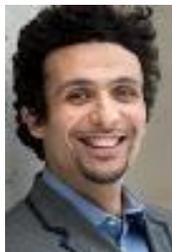

**Mehdi Bennis** (S'07-AM'08-SM'15) received his M.Sc. degree in Electrical Engineering jointly from the EPFL, Switzerland and the Eurecom Institute, France in 2002. From 2002 to 2004, he worked as a research engineer at IMRA-EUROPE investigating adaptive equalization algorithms for mobile digital TV. In 2004, he joined the Centre for Wireless Communications (CWC) at the University of Oulu, Finland as a research scientist. In 2008, he was a visiting researcher at the Alcatel-Lucent chair on flexible radio, SUPELEC. He obtained his Ph.D. in December 2009 on spectrum sharing for future mobile cellular systems. Currently, Dr. Bennis is an Associate Professor at the University of Oulu and Academy of Finland research fellow. His main research interests are in radio resource management, heterogeneous networks, game theory and machine learning in 5G networks and beyond. He has co-authored one book and published more than 100 research papers in international conferences, journals, and book chapters. He was the recipient of the prestigious 2015 Fred W. Ellersick Prize from

**IEEE COMSOC MMTC Communications - Frontiers**

the IEEE Communications Society, the 2016 Best Tutorial Prize from the IEEE Communications Society and the 2017 EURASIP Best Paper Award for the Journal of Wireless Communications and Networks. From 2015 to 2017 Dr. Bennis is currently an editor for the IEEE Transactions on Communications.